\documentclass[prd,showpacs,nofootinbib,preprintnumbers]{revtex4}
\usepackage[a4paper,left=20mm,right=20mm,top=25mm,bottom=25mm]{geometry}
\usepackage{color}
\usepackage{amsmath}
\usepackage{amsfonts}
\usepackage{graphicx}
\usepackage{dcolumn}
\usepackage{hyperref}

\begin{document}

\title{Cosmological dynamics of magnetic Bianchi I in viable $f(R)$ models of gravity}

\author{Xuyang Liu}
\affiliation{School of Mathematics and Physics, Bohai University, Liaoning, 121013, China}

\author{Phongpichit Channuie}
\affiliation{School of Science, Walailak University, Nakhon Si Thammarat, 80160 Thailand}

\author{Daris Samart}\email{daris.sa@rmuti.ac.th}
\affiliation{Department of Applied Physics, Faculty of Sciences and Liberal arts, Rajamangala University of Technology Isan, Nakhon Ratchasima, 30000, Thailand}
\affiliation{Center Of Excellent in High Energy Physics \& Astrophysics, Suranaree University of Technology, Nakhon Ratchasima, 30000, Thailand}

\date{\today}

\vskip 1pc
\begin{abstract}
Standard dynamical system analysis of Einstein-Maxwell equation in $f(R)$
theories is considered in this work. We investigate
cosmological dynamics of a uniform magnetic field in the Orthogonal Spatially Homogeneous (OSH)
Bianchi I universe with viable $f(R)$ models of gravity. In this work, the $f(R) = R -\alpha R^n$ and $f(R) = \left( R^b - \Lambda\right)^c$ models are examined by using our dynamical system analysis. Our results show that both of two $f(R)$ models have a viable cosmological consequence identical to the analysis present in Ref.\cite{Amendola:2007nt} for the FLRW background. Contrary to Ref.\cite{Amendola:2007nt}, we discover in our models that there is an additional anisotropic and non-zero cosmological magnetic fields fixed point emerging before the present of the standard matter epoch. This means that the universe has initially isotropic stage with the intermediated epoch as the anisotropic background and it ends up with the isotropic late-time acceleration. The primordial magnetic fields play a crucial role of the shear evolutions obtained from these two models which have the same scaling of the cosmic time as $\sigma\sim t^{-\frac13}$, instead of $\sigma\sim t^{-1}$ for the absence of the primordial magnetic cases.
\end{abstract}


\maketitle \vskip 1pc

\section{Introduction} \label{Intro}
An alternative way of explaining the accelerating expansion of the universe
\cite{SNa,CMB} or the Dark Energy (DE) problem \cite{CC} (for review see
\cite{Copeland:2006wr}) is to modify the Einstein's theory of gravity (General Relativity (GR)) as the source of accelerating behavior of the universe (for
review see \cite{reviewfR})\,. The simple versions of such modification, i.e. $f(R)$ gravity, were purposed by Capozziello et al. \cite{Capozziello:2003tk}
and Carroll et al. \cite{carrollmodel}\,. The main idea of modified
gravity is that, on the one hand, one considers gravitational Lagrangian as a function of
the Ricci scalar, i.e. $\mathcal{L}=\sqrt{-g}f(R)$ \cite{fR lagran}\,. On the other hand,
the inverse power of the Ricci scalar ($1/R$) is placed into the Einstein-Hilbert 
action which gives a deviation from GR at small curvature
and causes the present acceleration of the universe at very large scale. This
happens either with de Sitter and anti-de Sitter solutions in the
vacuum case which provides a purely gravitation driving the acceleration universe. The modified $f(R)$ gravity gives good explanation for the cosmic acceleration without introducing the dark energy component that implies from the cosmological data \cite{Cosmodata}\,. In addition, the modified $f(R)$ gravity also has been shown that it can be derived from string/M-theory \cite{Nojiri:2003rz}\,. 
However, $f(R)$ gravity encounters a number of problems, for example, the unstable problem of the scalar degree of freedom, especially in the $1/R$ model, \cite{Chiba:2003ir}: incompatible with the constraints of local gravity \cite{Chiba:2003ir,Olmo:2005}: the instability of cosmological perturbations in the large scale structure \cite{Carroll:2006jn} and the wrong sequence of the universe's evolution \cite{Amendola:2006kh}. 
The necessary conditions in both of local gravity and cosmological observations for viable $f(R)$ and its detailed implications are explicitly demonstrated and given in Refs.\cite{viablefR,Amendola:2006we,Amendola:2007nt,Starobinsky:2007hu,Hu:2007nk} and see Ref.\cite{reviewfR} for review.

The dynamics of anisotropic models with $f(R)$ gravity have been less studied compared with their isotropic Freidmann-Lemaitre-Robertson-Walker (FLRW) counterparts. As a result, it is not known how the behavior of the shear is modified in these theories of gravity. The dynamical systems methods for analysing the qualitative properties of cosmological models have proven very useful. It has been successfully used to study and to understand a number of cosmological models such as the standard GR cosmology \cite{Coley:2003mj}, the scalar fields models of dark energy \cite{Copeland:1997et}, the scalar-tensor theories of gravity \cite{Carloni:2007eu} and the brane-world models \cite{Coley:2001ab}. Moreover, the cosmological dynamics of $f(R)$ gravity was extensively studied in \cite{carrollmodel,dynamicsfR,Carloni:2004kp,Carloni:2007,Carloni:2007br,DSAiso,Leon:2013bra} by using the dynamical system analysis frameworks in homogeneous and isotropic universe (a.k.a. the FLRW model) and in the less anisotropic counterpart (Bianchi types and the others) \cite{Goheer:2007wu,Leach:2006br,Leach:2007ss,DSAaniso}. However, a huge number of $f(R)$ models in the FLRW universe is ruled out by the corrected cosmological expansion sequences \cite{Amendola:2006we}. A few models survives from both cosmological and local gravity constraints. Those models are called viable $f(R)$ models of gravity.
The interesting models among them are $f(R) = R - \alpha R^n$ and  $f(R) = (R^b - \Lambda)^c$ where they were purposed by Refs. \cite{Amendola:2006we} and \cite{Amendola:2007nt} respectively.
An extensive analysis of these viable $f(R)$ models in the anisotropic universe should be a very worth study to quantify some interesting features in this scenario of the $f(R)$ theories. Therefore, we will perform the dynamical system study for the cosmological solutions of $f(R)$ gravity in anisotropic Bianchi I universe with the existence of the uniform magnetic fields in this work.

The cosmological magnetic field is one of yet unsolved problems in
cosmology. The magnetic fields seem observationally emerge at large
scale of the universe \cite{Kronberg:1993vk}.
There are many theoretical explanations to disentangle the origin of primordial cosmological
magnetic field. One of the compelling explanations is that the cosmological magnetic field has a
primordial origin and this idea was purposed by Ref.\cite{zeldovich:1970}\,. The explanation is that it has been created from the Big Bang like all
matters populating the universe. From this assumption, we therefore include the primordial magnetic fields into the energy-momentum tensor
in the RHS of the Einstein field equation directly. This idea inspired us trying to understand
its evolution by finding the exact solutions
\cite{thorne:1967,jacob:1969,exact:magnetic} in the wide classes of
the Bianchi universe and its existences via the cosmological
perturbation theory approach (see for the Bianchi I background \cite{Tsagas:1999tu} and for a review \cite{Barrow:2006ch}). The cosmological magnetic fields will
naturally appear in the universe when the anisotropic cosmological models are taken
into account. More importantly, the (primordial) cosmological magnetic fields also might play some roles on the cosmic microwave background (CMB) radiation
and might be relics of the existence of the magnetic fields from the early universe.

Previously, the cosmological magnetic fields have been studied in the homogenous
anisotropic universe (Bianchi models) context. It was mentioned in Refs.\cite{thorne:1967} that a universe with a primordial magnetic field is necessary anisotropy . The exact solutions of magnetic fields in many classes of the Bianchi models were discovered \cite{thorne:1967,jacob:1969,exact:magnetic}. The first significant study of cosmological
magnetism in the dynamical system approach was performed by the authors of Ref.\cite{Collins:1971gc}\,. The authors of \cite{LeBlanc:1994qm} was
systematically performed the standard technique of dynamical system
of the magnetic fields evolutions with the perfect fluids in the
Bianchi IV$_0$ with the orthogonal frame formalisms and Hubble
normalization variables. Later, there were many works on the
dynamical system approach to study the magnetic fields in several classes
of the Bianchi models \cite{Clarkson:2001fd,Horwood:2003dg,LeBlanc:1995qc} (see also
reference therein) with the GR and the brane-world cosmology \cite{Barrow:2001bp}.

In this work, we will perform the standard dynamical system analysis on
the dynamics of the Orthogonally Spatially Homogenous (OSH) Bianchi I model
in $f(R) = R - \alpha R^n$ and  $f(R) = (R^b - \Lambda)^c$ models of gravity with the existence of the primordial magnetic fields and the standard barotropic perfect fluid matter ($p_m=w\rho_m$) for $w=1$.
The two selected viable $f(R)$ models in this work have advantages in both local gravity constraints and cosmological viabilities which are demonstrated in Refs.\cite{Amendola:2006we,Amendola:2007nt} for the FLRW case.  The Bianchi I is reasonable to be examined because such models are anisotropic generalization of the flat FLRW model and its mathematical simplicity as well. The geometrical property of the spacetime in the Bianchi I, in this work, is assumed to exhibit a property of \lq\lq Locally Rotational Symmetry\rq\rq (LRS) \cite{Leach:2006br}. The LRS is rotational symmetry with a preferred spatial direction of the space-like hypersurface. Physical quantities and also observations are rotationally invariant about this direction \cite{Wainwright:1989nm,Ellis:1968vb,vanElst:1996dr}. Here the $x$-direction is chosen to be the axis of this symmetry. Therefore, we impose the uniform magnetic field aligned along the axis of the LRS (in the $x$-direction). The aim of this work is also to study the cosmological dynamics of anisotropic universe with the
magnetic fields in $f(R)$ gravity via the dynamical system technique.
The $f(R)$ gravity might have some interesting features on the anisotropies in the shear evolutions of the early universe till the present observed universe.
The results from this study might help to understand more about the alternative effects of the viable $f(R)$ gravity DE models
on the small anisotropic effect and contributions of magnetic fields in CMB and its evolution epoch of the universe.
In addition, this work is the first systematic dynamical system analysis of the magnetic Bianchi I in $f(R)$ models of gravity
which has never been studied.

We outline this work as follows: In section 2 we setup the evolution equations of
the $f(R)$ gravity in OSH $1+3$ covariant approach. In the next section, the dynamical system analysis of the magnetic Bianchi I in the in $f(R) = R - \alpha R^n$ and  $f(R) = (R^b - \Lambda)^c$ models is investigated. In section 4, we discuss the cosmological implications stemming from our study. In the last section, we give the conclusion in this work.  Unless otherwise specified, we will use natural units ($c=8\pi G=\hbar =1$) and all conventions used in the present work are adopted from Refs.\cite{Goheer:2007wu,Leach:2006br,Leach:2007ss}.

\section{Evolution equations of $f(R)$ gravity in 1+3 covariant formalisms}
\label{sec:eveq}
In this section, we will briefly give a crucial setup for the $f(R)$ gravity in terms of the 1+3 covariant formalisms. We mainly follow the notations and conventions from Refs.\cite{Rippl:1995bg,Leach:2006br} and its detailed formalisms there in.
\subsection{The Einstein field equation in $f(R)$ gravity}
We begin at the gravitational action of $f(R)$ gravity in the following form
\begin{equation}\label{action}
S=\int \sqrt{-g}f(R) d^4x+\int \mathcal{L}_m d^4x
\end{equation}
where $\mathcal{L}_m$ is matter fields lagrangian density. Varying
above action with respect to metric $g^{ab}$\,, we obtain the
Einstein field equation of $f(R)$ as
\begin{equation}\label{EFE}
F(R)R_{ab}-\frac{1}{2}\,f(R)g_{ab}+g_{ab}\nabla_c\nabla^c
F(R)-\nabla_a\nabla_b F(R)=T_{ab}\,.
\end{equation}
where $F\equiv \partial f/\partial R$ and $T_{ab}\equiv
(2/\sqrt{-g})\delta \mathcal{L}_m/\delta g^{ab}$\, and the Latin indices run from $a,\,b,\,c\,= 0,...,3$. After some manipulations, the Ricci tensor takes form \cite{Leach:2006br,Rippl:1995bg},
\begin{equation}
\label{eq:einstScTneff}
R_{ab}=\frac{1}{F}\left(\frac{1}{2}g_{ab}f-(g_{ab}g^{cd}-g_a^cg_b^d)S_{cd}+T_{ab}\right)
\end{equation}
where $S_{ab}=\nabla_a\nabla_b F$\,. The energy-momentum tensor
$T_{ab}$ is defined by
\begin{equation}
T_{ab}=\rho u_au_b+ph_{ab}+q_au_b+q_bu_a+\pi_{ab}\,.
\end{equation}
Here $h_{ab}=g_{ab}+u_au_b$ is induce metric which associated with spatial hypersurface and $u^a=(\,1\,,~0\,,~0\,,~0\,)$ is four-velocity which orthogonal to $h_{ab}$ ($h_{ab}u^a=0$)\,, $q_a$ is the energy flux ($q_au^a=0$) and $\pi_{ab}$ is the symmetric trace-free anisotropic pressure, all relative to $u^a$\, ($\pi_{a}^a=0\,, \pi_{ab}u^a=0$)\,, \cite{Leach:2006br}.
The energy momentum tensor in this work can be decomposed into two parts as
\begin{equation}
T_{ab}=T_{ab}^{PF}+T_{ab}^{EM}
\end{equation}
\indent where $T_{ab}^{PF}$ is the energy-momentum tensor of the perfect fluid,
given by
\begin{equation}
T_{ab}^{PF}=\rho_m u_au_b+p_mh_{ab}\,.
\end{equation}
\indent $T_{ab}^{EM}$ is the energy-momentum tensor of Maxwell
field, given by
\begin{eqnarray}
T_{ab}^{EM}&=&F_{ac}F^{c}_{~b}-\frac{1}{4}g_{ab}F_{cd}F^{cd}
\end{eqnarray}
the field strength $F_{ab}$ is also defined by
\begin{equation}
F_{ab}=\frac{1}{2}\,u_{[a}E_{b]}+\eta_{abcd}H^cu^d
\end{equation}
where $E_a$ and $H_a$ are electric and magnetic fields respectively.
We will consider the energy-momentum tensor of Maxwell field in the pure magnetic
case. The uniform magnetic fields are aligned in $x$-direction. This means the magnetic fields have
component as $H_a=(\,0\,,~\tilde H\,,~0\,,~0\,)$\, \cite{LeBlanc:1994qm}\,. One can write the energy-momentum tensor of the Maxwell fields analogous with the perfect fluid as \cite{Barrow:2001bp}
\begin{equation}
T_{ab}^{EM}=\rho_{EM}u_au_b+p_{EM}h_{ab}+\pi_{ab}
\end{equation}
where $\rho_{EM}=\frac{1}{2}\tilde H^2$\,, $p_{EM}=\frac{1}{6}\tilde H^2$\, and $\pi_{ab}$ is given by
\begin{equation}
\pi_{ab}=\frac{1}{3}H^2h_{ab}-H_aH_b \,,
\end{equation}where $H^2\equiv H_aH^a = \tilde H^2 $ is the magnitude of magnetic fields.\\
\indent One notes that the energy-momentum tensor of Maxwell field is also
trace-free ($g^{ab}T_{ab}^{EM}=T^{EM}=0$)\,. Then the energy density
$\rho$ and pressure $p$ can be decomposed to the fluid matter and the magnetic
parts as
\begin{equation}
\rho = \rho_{PF} + \rho_{EM}\,,\quad \rho_{PF}=\rho_m\,,\quad \rho_{EM}=\frac{1}{2}\,\tilde H^2\,,\qquad\quad p = p_{PF} + p_{EM}\,,\quad p_{PF}=p_m\,,\quad p_{EM}=\frac{1}{6}\tilde H^2\,.
\end{equation}
Here we consider the energy-momentum tensor of matter part as a
standard perfect fluid (the equation of states for perfect fluid and EM parts take form
$w=p_m/\rho_m$ and $w_{EM}=1/3$ respectively.)\,.

\subsection{Propagation equations of Kinematical quantities in $f(R)$ gravity}
In the next step, we will set up the propagation equations for $f(R)$
gravity in OSH $1+3$ covariant formalism. The OSH formalism is used to describe the fluid velocity time-like vector fields which orthogonalize
to the spatial vector (triad) fields. Having assumed the LRS of the spacetime, here the spatial vector fields span on the space-like hypersurface with one preferred spatial direction and they are invariant under the rotation of the preferred spatial axis (the $x$-direction) \cite{Wainwright:1989nm,Ellis:1968vb,vanElst:1996dr}.
The main results have been done by \cite{Leach:2006br,Rippl:1995bg,Maartens:1994pb}. Using Eq.(\ref{eq:einstScTneff})\,, we can split $R_{ab}$ in the following forms as
\cite{Leach:2006br,Rippl:1995bg}
\begin{eqnarray}
R&=&F^{-1}(T+2f-3S)\\
R_{ab}u^au^b&=&F^{-1}(T_{ab}u^au^b-\frac{1}{2}f+h^{ab}S_{ab})\\
R_{ab}u^ah_c^b&=&F^{-1}(S_{ab}u^ah_c^b-q_c)\\
R_{ab}h_c^ah_d^b&=&F^{-1}\left(\pi_{cd}-(p+\frac{1}{2}f+S)h_{cd}+S_{ab}h_c^ah_d^b\right)\,,
\end{eqnarray} and for the $S_{ab}$,
\begin{eqnarray}
S&=&-F'(\ddot{R}+\Theta\dot{R})-F''\dot{R}^2\\
S_{ab}u^au^b&=&F'\ddot{R}+F''\dot{R}^2\\
S_{ab}h^{ab}&=&-F'\Theta\dot{R}\,.
\end{eqnarray}
According to Refs.\cite{Leach:2006br} and \cite{Goheer:2007wu}, they were
explicitly demonstrated the $1+3$ covariant analysis of Bianchi 
I universe in the $f(R)$ gravities, the Raychaudhuri equation is
written by
\begin{eqnarray}\label{raychua}
&& \dot{\Theta}+\frac{1}{3}\Theta^2+2\sigma^2+\frac{1}{F}\left(\rho-\frac{1}{2}f+h^{ab}S_{ab}\right) =
\nonumber\\
&&
\dot{\Theta}+\frac{1}{3}\Theta^2+2\sigma^2+\frac{1}{F}\left(\rho-\frac{1}{2}f - F'\Theta\dot{R} \right)=0\,,
\end{eqnarray}
and the Friedmann equation (the first integral of the above equation) is given by
\begin{eqnarray}\label{friedmann}
&&\frac13\Theta^2 - \sigma^2 - \frac{1}{F}\left(\rho + 3p + f - 3S + 2h^{ab}S_{ab} \right) =
\nonumber\\
&&\frac13\Theta^2 - \sigma^2 - \frac{1}{F}\left(\rho + \frac{1}{2}(FR - f) - F'\Theta\dot{R} \right) = 0
\end{eqnarray}
The shear propagation equation is given by
\begin{equation}\label{sheardot}
\dot{\sigma}_{ab}+\Theta\sigma_{ab}=\frac{1}{F}\left(
\pi_{ab}-F'\dot{R}\,\sigma_{ab}\right),
\end{equation}
where $\Theta\equiv \Theta_{ab}h^{ab}$ is the rate of volume
expansion parameter (Hubble parameter), $\sigma^2\equiv
\frac{1}{2}\sigma_{ab}\sigma^{ab}$ is magnitude of the shear tensor
$\sigma_{ab}$ ($\sigma_{ab}=\Theta_{ab}-\frac{1}{3}h_{ab}\Theta$\,,
$\sigma_a^a=0$ and $\sigma_{ab}u^a=0$)\,. The tetrad field is
decomposed into the orthonormal frame \cite{vanElst:1996dr}\,. We
restrict that the magnetic field is aligned along the shear
eigenvector as were done in \cite{LeBlanc:1994qm,Clarkson:2001fd}\,,
then the shear tensor simultaneously diagonalize as\footnote{Alternatively, by using the LRS Bianchi I line element, $ds^2 = -dt^2 + X(t)^2dx^2 + Y(t)^2\big(dy^2 + dz^2\big)$, one can show that $\sigma_{11} = \Theta_{11} - h_{11}\Theta = \frac23\left( \frac{\dot X}{X} - \frac{\dot Y}{Y}\right)$, $\sigma_{22} = \Theta_{22} - h_{22}\Theta = -\frac13\left( \frac{\dot X}{X} - \frac{\dot Y}{Y}\right)$ and $\sigma_{33} = \Theta_{33} - h_{33}\Theta = -\frac13\left( \frac{\dot X}{X} - \frac{\dot Y}{Y}\right)$ where $\Theta = \frac{\dot X}{X} + 2 \frac{\dot Y}{Y}$, $\Theta_{11} = \frac{\dot X}{X}$, $\Theta_{22} = \frac{\dot Y}{Y}$ and $\Theta_{33} = \frac{\dot Y}{Y}$. From such results, we obtain $\sigma_{11} = \sigma_1 = -(\sigma_2 + \sigma_3)$.}
\begin{equation}
\sigma_{ab}=\text{diag}(\,\sigma_{11}\,,~\sigma_{22}\,,~\sigma_{33}\,)
\equiv \text{diag}(\,\sigma_{1}\,,~\sigma_{2}\,,~\sigma_{3}\,)\,.
\end{equation}
Therefore the shear propagation can be written in the following form
\begin{eqnarray}\label{sheardia}
\dot{\sigma}_{a}+\Theta\sigma_{a}=\frac{1}{F}
\left(\pi_{a}-F'\dot{R}\,\sigma_{a}\right),
\end{eqnarray}
where $\pi_a \equiv \pi_{aa}$ and $\pi_{aa}$ is the diagonal elements of $\pi_{ab}$ tensor\,.
By using the conservation of
energy-momentum tensor with source-free of Maxwell field in the the Bianchi I scenario, the
propagation of matter parts is given by \cite{LeBlanc:1994qm}
\begin{eqnarray}\label{rhodot}
\dot{\rho_m}&=&-(1+w)\rho_m\Theta,
\end{eqnarray}
\begin{eqnarray}\label{hdot}
\dot{\tilde H}&=&-\frac{2}{3}\Theta \tilde H + \sigma_{11}\tilde H = -\frac{2}{3}\Theta \tilde H - 2(\sigma_2 + \sigma_3)\tilde H\,.
\end{eqnarray}
We close this section by introducing another helpful equation. With help of the Raychaudhuri and Freidmann equations, we come up with the following crucial relation:
\begin{eqnarray}\label{R-thetadot}
R &=&2\dot{\Theta} + \frac{4}{3}\Theta^2 + 2\sigma^2\,.
\end{eqnarray}%
\subsection{The autonomous system}
\label{sec:analyze}
According to Ref.\cite{Amendola:2006we}, we introduce new dimensionless variables
as follows:
\begin{eqnarray}\label{dlessvar}
\Sigma &=&\frac32\frac{\sigma_2 + \sigma_3}{\Theta}\,,\qquad\quad M = \sqrt{\frac{3}{2F}}\frac{\tilde H}{\Theta}\,\,,
\nonumber\\
x_1 &=& -\frac{3}{\Theta}\frac{F'\dot{R}}{F}\,,\qquad\quad~\,  x_2 = -\frac{3}{2}\frac{f}{F \Theta^2}\,,
\nonumber\\
x_3 &=& \frac{3}{2}\frac{R}{\Theta^2}\,,\qquad\qquad\quad~~ z = \frac{3}{\Theta^2}\frac{\rho_m}{F}\,.
\end{eqnarray}
One notes that\footnote{We can demonstrate that $\left( \frac32\frac{\sigma_2 + \sigma_3}{\Theta} \right)^2 = 3\frac{\sigma^2}{\Theta^2} = \frac32\frac{\sigma_1^2+\sigma_2^2+
\sigma_3^2}{\Theta^2}$ is valid by using the relation $\sigma_{1}=\frac23\left( \frac{\dot X}{X} - \frac{\dot Y}{Y}\right)$ and $\sigma_{2}=\sigma_3=-\frac13\left( \frac{\dot X}{X} - \frac{\dot Y}{Y}\right)$ } $\Sigma^2 = \left( \frac32\frac{\sigma_2 + \sigma_3}{\Theta} \right)^2 = 3\frac{\sigma^2}{\Theta^2}$. The constraint equation written in terms of the new variables obeys
\begin{eqnarray}\label{dlessconstr}
1-\Sigma^2-x_1-x_2-x_3-z-M^2=0\,.
\end{eqnarray}
Using the time propagation equations of the kinematical quantities
in the previous section, then the equations of autonomous system are given
\begin{eqnarray}\label{autonomous}
\frac{d\Sigma}{d\tau} &=& x_1\Sigma - 3\Sigma + \Sigma\left( 2 - x_3 + \Sigma^2 \right)
+ 2\left( 1- \Sigma^2 -x_1 - x_2 - x_3 - z \right),
\nonumber\\
\frac{d x_1}{d\tau} &=& x_1^2 - 3x_1 - 4x_2 - 2x_3 - (1-3w)z + x_1\left( 2 - x_3 + \Sigma^2 \right),
\nonumber\\
\frac{d x_2}{d\tau} &=& x_1x_2 + \frac{x_1x_3}{m} + 2x_2\left( 2 - x_3 + \Sigma^2 \right),
\nonumber\\
\frac{d x_3}{d\tau}&=& -\frac{x_1x_3}{m} + 2x_3\left( 2 - x_3 + \Sigma^2 \right),
\nonumber\\
\frac{d z}{d\tau}&=& x_1z - 3(1+w)z +2z\left( 2 - x_3 + \Sigma^2 \right),
\end{eqnarray}
where $m$ is the parameter which is very useful to study viable models of $f(R)$ gravity and it is defined by \cite{Amendola:2006we}
\begin{equation}\label{decela}
m = \frac{RF'}{F}\,.
\end{equation}
We note that the derivative with
respect to the logarithm time scale is defined by
$\frac{d~}{d\tau}=\frac{3}{\Theta}\frac{d~}{dt}\,.$ In addition, one observes that by ignoring the $\Sigma$ and $M$ variables, the autonomous system will be identical to the FLRW case that have been done in \cite{Amendola:2006we}. For the general
case of the evolution phase of the universe, one can be described by the sign of the rate of volume expansion $\epsilon=\pm 1$\,, where
$\epsilon\equiv |\Theta|/\Theta$\,. $\epsilon=1$ for the expanding
phase and $\epsilon=-1$ for the collapsing phase, more detail
discussion in this issue see
\cite{Carloni:2007,Leach:2007ss,Goheer:2007wu}. In this work, we
have focused our study in the future evolution of the expanding
phase ($\epsilon=1$) of the universe only. The auxiliary of the
autonomous system (evolution of magnetic term) is 
\begin{equation}\label{Hdyn}
\frac{d M}{d\tau}= \frac{x_1M}{2} -2M - 2\Sigma M + M\left( 2 - x_3 + \Sigma^2 \right)\,.
\end{equation}
The matter density and curvature density parameters of the
universe are defined by $\Omega_m\equiv z$ and $\Omega_c\equiv x_1 + x_2 + x_3$,
respectively. In this work we consider $\Omega_c$ as dark energy density parameter.\\
\indent This section we note that \lq\lq prime\rq\rq denotes derivatives with
respect to the logarithm time scale as
$\frac{d~}{d\tau}=\frac{3}{\Theta}\frac{d~}{dt}\,.$ In the general
case of the evolution phase of the universe, it can be described by the
sign of the rate of volume expansion $\epsilon=\pm 1$\,, where
$\epsilon\equiv |\Theta|/\Theta$\,. $\epsilon=1$ for the expanding
phase and $\epsilon=-1$ for the collapsing phase, more detailed
discussion in this issue see
\cite{Carloni:2007,Leach:2007ss,Goheer:2007wu}. We also introduce the effective equation of state \cite{Amendola:2006we},
\begin{eqnarray}\label{weff}
w_{\rm eff} &=& -1 - 2\frac{\dot \Theta}{\Theta^2},
\end{eqnarray}
which is a useful parameter in this study. Using the definitions in Eq. (\ref{dlessvar}), one rewrites the effective equation of states in terms
of the dimensionless variables as
\begin{eqnarray}\label{dlessweff}
w_{\rm eff} &=& -1 - \frac{2}{3}\left( x_3 - \Sigma^2 - 2 \right).
\end{eqnarray}
More importantly, the associated solutions for each fixed points can be obtained by using Eqs. (\ref{R-thetadot}) and (\ref{dlessvar}) to yield
\begin{eqnarray} \label{dotH}
\dot \Theta = \left( x_{3,(i)} - \Sigma_{(i)}^2 - 2 \right)\frac{\Theta^2}{3} \,,
\end{eqnarray}
where $x_{3,(i)}$ and $\Sigma_{(i)}$ are the associated ``$\,i\,$" fixed points.
Performing the integration, one gets solutions of the scale factor for the associated fixed points
\begin{eqnarray} \label{at}
a(t) &=& a_0\left(t - t_0\right)^\beta\,,
\nonumber\\
\beta &=& \frac{1}{ 2 + \Sigma_{(i)}^2 - x_{3,(i)} }\,.
\end{eqnarray}

In order to see dynamical features of the anisotropic universe, one should consider the shear evolution in the model. With help of Eqs.(\ref{sheardia}) and (\ref{dlessvar}), we obtain the shear evolution equation in terms of the dimensionless variables as
\begin{eqnarray} \label{shear}
\frac{\dot \sigma}{\sigma} = -\eta\Theta,\qquad \eta = \frac13\left( 3 - 2\frac{M^2_{(i)}}{\Sigma_{(i)}} - x_{1,(i)}\right).
\end{eqnarray}
Contrary to previous studies on the $R^{n}$ gravity in Ref.\cite{Leach:2006br}, our parameter $\eta$ depends on the magnetic field, $M$. This means has the magnetic fields play important role on the shear evolution. Integrating out Eq.(\ref{shear}), we find
\begin{eqnarray}\label{exactshear}
\sigma = \sigma_0 a^{-\eta} = \sigma_0 a_0^{-\eta} (t-t_0)^{-\beta\eta}\,.
\end{eqnarray}
The exact solution of the shear evolution is very useful for understanding the behavior of the anisotropic effect in the universe. As discussed in Ref. \cite{Leach:2006br}, from the above equation the shear evolution for all points in the phase space that lie on the line $\eta\equiv\left(3 +2M^2_{(i)}/\Sigma_{(i)}+x_{1,(i)}\right)/3 =1$ is the same as in the GR case. In order to deviate from the standard GR, the shear will dissipate faster than that in GR when $\dot{\sigma}/\sigma < -\Theta$, that is all points that lie in the region $\left(3 +2M^2_{(i)}/\Sigma_{(i)}+x_{1,(i)}\right)/3 > 1$. This is called the fast shear dissipation (FSD) regime \cite{Leach:2006br}. When $\dot{\sigma}/\sigma > -\Theta$ and for all points in the region $\left(3 +2M^2_{(i)}/\Sigma_{(i)}+x_{1,(i)}\right)/3 < 1$, the shear will dissipate slower than that in GR. This is named the slow shear dissipation (SSD) regime \cite{Leach:2006br}. Notice that, however, the higher order terms of the $f(R)$ gravity models, e.g. see Refs.\cite{Nojiri:2014jqa}, can play the same role as of the magnetic field investigated in this work.

In addition, exact solution of the magnetic fields are also obtained in terms of dimensionless variables. Using Eqs. (\ref{hdot}) and (\ref{dlessvar}), one gets,
\begin{eqnarray}\label{exactmagnetic}
\tilde H = \tilde H_0 a^{-\kappa} = \tilde H_0 a_0^{-\kappa} (t -t_0)^{-\beta \kappa}\,,\qquad \kappa = \frac{2}{3}\left( 1 + 2\Sigma_{(i)}\right).
\end{eqnarray}
In order to see how the shear parameter and magnetic fields evolve in cosmic time, we will substitute non-zero shear fixed points into the exact solutions of the shear and the magnetic fields evolutions, Eqs. (\ref{exactshear}) and (\ref{exactmagnetic}).
After outlining the autonomous system of the magnetic Bianchi I universe with the generic $f(R)$ gravity and exact solutions in terms of dimensionless variables, we will consider such system of differential equations by using the standard dynamical system approach in next section.

\section{Dynamics of magnetic Bianchi I universe in $f(R)$ models of gravity}
This section is devoted to provide the dynamical system analysis for the $f(R)$ models of gravity. We will define the dimensionless variables from the Friedmann equation given in Eq.(\ref{friedmann}) in the previous section and use these variables to setup a autonomous system of first-order non-linear differential equations. Next we will determine all fixed points of the autonomous system and analyze their stabilities for each of them. At the end of this section, the cosmological implications of the magnetic Bianchi I in $f(R)$ models will be discussed in accord with the fixed points and its stabilities.
\subsection{Dynamical system of the $f(R) = R - \alpha R^n$ gravity}
We start with the $f(R) = R - \alpha R^n$ gravity. This model has been studied extensively in several aspects. and it was shown that the model will be the viable $f(R)$ DE if $\alpha > 0$ and $0<n<1$ \cite{Amendola:2006we}. Especially, the standard dynamical system method is used to analyze in the FLRW counterpart \cite{Carloni:2007br}. For the anisotropic cases, it was studied in the Kantowski-Sach metric \cite{Leon:2013bra}. Here we will consider this model in the Bianchi I universe with the existence of the primordial magnetic fields. The $f(R) = R - \alpha R^n$ model has the $m$ function which can be practically written in terms of a variable $r$ as
\begin{eqnarray}
m = \frac{n(1+r)}{r}  \,,
\end{eqnarray}
where $r \equiv x_3/x_2$. Substituting the $m$ function into the autonomous system in Eq.(\ref{autonomous}), one obtains explicit dynamical system for the $f(R) = R - \alpha R^n$ gravity. The dynamical system of this model is given as follows:
\begin{eqnarray}\label{autonomous:model1}
\frac{d\Sigma}{d\tau} &=& x_1\Sigma - 3\Sigma + \Sigma\left( 2 - x_3 + \Sigma^2 \right)
+ 2\left( 1- \Sigma^2 -x_1 - x_2 - x_3 - z \right),
\nonumber\\
\frac{d x_1}{d\tau} &=& x_1^2 - 3x_1 - 4x_2 - 2x_3 - z + x_1\left( 2 - x_3 + \Sigma^2 \right),
\nonumber\\
\frac{d x_2}{d\tau} &=& x_1x_2 + \frac{x_1x_3^2}{n(x_2 + x_3)} + 2x_2\left( 2 - x_3 + \Sigma^2 \right),
\nonumber\\
\frac{d x_3}{d\tau}&=& -\frac{x_1x_3^2}{n(x_2 + x_3)} + 2x_3\left( 2 - x_3 + \Sigma^2 \right),
\nonumber\\
\frac{d z}{d\tau}&=& x_1z - 3z +2z\left( 2 - x_3 + \Sigma^2 \right).
\end{eqnarray}

We will separately study the fixed points, their stabilities, the shear and the magnetic evolutions below.

\subsubsection{Fixed points and their stabilities}
In what follows, we will consider the properties of each point in turn. There are 4 physical fixed points from the autonomous system of the $f(R) = R - \alpha R^n$ gravity. We will classify into two cases: isotropic and anisotropic solutions, and the physically associated fixed points of this model are given below.
\begin{center}
\it{Isotropic solutions}
\end{center}
\begin{itemize}
\item ${\rm (1)}\,\,P_{1}^{(1)}$\,: de-Sitter fixed point

In this case, we obtain the fixed point:
\begin{eqnarray}
\Sigma = 0,~x_1= 0,~x_2= -1,~x_3= 2,~z= 0.
\end{eqnarray}
Since $w_{{\rm eff}}=-1$, the point $P_{1}^{(1)}$ corresponds to de-Sitter
solutions ($\dot{\Theta}=0$) and has eigenvalues
\begin{equation}
\left\{-4,-3,-3,\frac{-3 n-\sqrt{n} \sqrt{-32+25 n}}{2 n},\frac{-3 n+\sqrt{n} \sqrt{-32+25 n}}{2 n}\right\}.\nonumber
\end{equation}
Hence $P_{1}^{(1)}$ is stable when $0\leq n<2$ and saddle for $n>2$. In this case, it is trivial to verify that,
\begin{equation}
a(t) = \exp(\lambda t), \quad \lambda = {\rm arbitrary~constant},\quad M^2=0\,.\label{m1con2}\nonumber
\end{equation}

\item $P_{2}^{(1)}$\,: standard matter-liked epoch fixed point

In this case, we obtain the fixed point:
\begin{eqnarray}
\Sigma = 0,~x_1= 3-\frac{3}{n},~x_2= \frac{3-4 n}{2 n^2},~x_3= 2-\frac{3}{2 n},~z = \frac{(13-8 n) n-3}{2 n^2}.
\end{eqnarray}
Employing Eqs.(\ref{dlessweff})-(\ref{at}), we obtain in this case
\begin{equation}
w_{\rm eff} = -1 +\frac1n,\qquad a(t) = a_0(t-t_0)^{\frac{2n}{3}},\quad M^2=0\,.\nonumber
\end{equation}
This point $P_{2}^{(1)}$ corresponds to saddle solutions and has eigenvalues
\begin{equation}
\left\{-1,-\frac{3}{2 n},\frac{3 (-1+n)}{n},N^{-},N^{+}\right\},\nonumber
\end{equation}
where
\begin{equation}
N^{\pm}\equiv \frac{3 n-3 n^2 \pm n \sqrt{81-498 n+1025 n^2-864 n^3+256 n^4}}{4 \left(-n^2+n^3\right)}.\nonumber
\end{equation}
The solutions are a saddle point for $\frac{1}{16} \left(13-\sqrt{73}\right)<n<3/4$. We note that this fixed point becomes the standard matter epoch if $n=1$.

\item $P_{3}^{(1)}$\,: curvature dominated fixed point

In this case, we obtain the fixed point:
\begin{eqnarray}
\Sigma = 0,~x_1= \frac{3}{2 n-1}-1,~x_2= \frac{6}{1-2 n}+\frac{1}{n-1},~x_3= \frac{3}{2 n-1}+\frac{1}{1-n}+2,~z= 0.
\end{eqnarray}
Employing Eqs.(\ref{dlessweff})-(\ref{at}), we obtain in this case
\begin{equation}
w_{\rm eff} = -1 +\frac{-10 n^2+13 n-1}{6 n^2-9 n+3},\qquad a(t) = a_0(t-t_0)^{-\frac{(n-1) (2 n-1)}{n-2}},\quad M^2=0\,.\nonumber
\end{equation}
This point $P_{3}^{(1)}$ has eigenvalues
\begin{equation}
\left\{\frac{5-4 n}{-1+n},\frac{5-4 n}{-1+n},-\frac{2 (-2+n)}{-1+2 n},-\frac{2 \left(2-8 n+5 n^2\right)}{1-3 n+2 n^2},\frac{-3+13 n-8 n^2}{1-3 n+2 n^2}\right\}.\nonumber
\end{equation}
Regarding the above values, we find for unstable fixed points
\begin{equation}
1<n\leq \frac{5}{4},\label{usp13}
\end{equation}
and for stable ones
\begin{equation}
n<\frac{1}{16} \left(13-\sqrt{73}\right)\lor n>2.\label{stp13}
\end{equation}

\begin{center}
\it{Anisotropic solutions}
\end{center}

\item $P_{4}^{(1)}$\,: Jacob magnetic-like (non-zero magnetic field with matter solution: the Jacobs magnetic field model in Bianchi I \cite{jacob:1969}) with curvature fixed point

In this case, we obtain the fixed point:
\begin{eqnarray}
\Sigma &=& -\frac{2 (n (5 n-8)+2)}{n (7 n-10)+4},~x_1= -\frac{12 (n-2) (n-1)}{n (7 n-10)+4},~x_2= -\frac{18 (n-1) (n (11 n-20)+8)}{(n (7 n-10)+4)^2},
\nonumber\\
x_3 &=& \frac{18 (n-1) n (n (11 n-20)+8)}{(n (7 n-10)+4)^2},~z= 0.
\end{eqnarray}
Employing Eqs.(\ref{dlessweff})-(\ref{at}), we obtain in this case
\begin{equation}
w_{\rm eff} = -1 +\frac{-245 n^4+1616 n^3-492 n^2+464 n+16}{3 \left(7 n^2+10 n+4\right)^2},\qquad a(t) = a_0(t-t_0)^{\frac{\left(7 n^2+10 n+4\right)^2}{6 \left(193 n^3+24 n^2+72 n+8\right)}}\,,\nonumber
\end{equation}
\begin{equation}
M^2=-\frac{3 \left(55 n^4-188 n^3+222 n^2-104 n+16\right)}{\left(7 n^2-10 n+4\right)^2}\,.\nonumber
\end{equation}
The existence of magnetic fields in this case satisfies
\begin{equation}
\frac{1}{5} \left(4-\sqrt{6}\right)<n<\frac{2}{11} \left(5-\sqrt{3}\right)\lor \frac{2}{11} \left(5+\sqrt{3}\right)<n<\frac{1}{5} \left(4+\sqrt{6}\right).\label{magp14}
\end{equation}
This point $P_{4}^{(1)}$ has eigenvalues
\begin{equation}
\left\{-\frac{12 \left(2-3 n+n^2\right)}{4-10 n+7 n^2},-\frac{3 \left(8-20 n+11 n^2\right)}{4-10 n+7 n^2},-\frac{3 \left(4-18 n+11 n^2\right)}{4-10 n+7 n^2},P,Q\right\}\nonumber
\end{equation}
where
\begin{equation}
P\equiv-\frac{3 \left(32-160 n+300 n^2-250 n^3+77 n^4+B\right)}{2 \left(4-10 n+7 n^2\right)^2}\nonumber
\end{equation}
and
\begin{equation}
Q\equiv\frac{3 \left(-32+160 n-300 n^2+250 n^3-77 n^4+B\right)}{2 \left(4-10 n+7 n^2\right)^2}\nonumber
\end{equation}
with $B=\sqrt{\left(4-10 n+7 n^2\right)^2 \left(320-1984 n+4128 n^2-3448 n^3+1001 n^4\right)}$.
\end{itemize}
Before going further to the next subsection, we will give some discussion towards these fixed points of the $f(R) = R - \alpha R^n$ model. we find  from this model of gravity that there are 4 fixed points for the isotropic and for the anisotropic cases. Interestingly, there is no Kasner fixed point contrary to that of the usual GR gravity. According to the fixed points in this model, it means no anisotropic singularity in this scenario. We classify the physical fixed points by considering the magnitude square of the magnetic fields which must be positive.

\subsubsection{The shear and magnetic fields evolutions}
We turn to consider the shear and magnetic fields evolutions for the $f(R) = R - \alpha R^n$ model. According to the existence of the 4 physical fixed points above. There is only one fixed point with non-zero shear and magnetic solution. To see how the shear and magnetic fields dissipate, we recall the exact solution of the shear evolution from Eq.(\ref{exactshear}) and substitute the anisotropic fixed point in the solution to yield
\begin{eqnarray}
\sigma = \sigma_0 a^{-\eta} = \sigma_0 a_0^{\left(\frac{2 n-4}{7 n^2-10 n+4}\right)} (t-t_0)^{-\frac{1}{3}}\,,\qquad \eta = \frac{4- 2 n}{7 n^2-10 n+4}\,.
\end{eqnarray}
In the above results from the shear evolution, we find the shear dissipation scale in cosmic time as $\sigma \sim t^{-\frac{1}{3}}$. The $\eta$ parameter can be classified into FSD and SSD as
\begin{eqnarray}\label{SSDFSD1}
&& \frac{4 - 2 n}{7 n^2-10 n+4} > 1,\quad \stackrel{\rm FSD}{\rightarrow} \quad 0<n<\frac{8}{7}\,,
\nonumber\\
&& \frac{4 - 2 n}{7 n^2-10 n+4} < 1,\quad \stackrel{\rm SSD}{\rightarrow} \quad n<0\lor n>\frac{8}{7}\,.
\end{eqnarray}
The exact solution of the magnetic fields is given by,
\begin{eqnarray}
\tilde H = \tilde H_0 a^{-\kappa} = \tilde H_0 a_0^{\left(\frac{26 n^2-44 n+8}{21 n^2-30 n+12}\right)} (t -t_0)^{\left(\frac{13 n^2-22 n+4}{18-9 n}\right)}\,,\qquad \kappa = -\frac{26 n^2-44 n+8}{21 n^2-30 n+12}.
\end{eqnarray}
We will see the numeric results of the shear and the magnetic fields evolutions in the section IV.
\subsection{Dynamical system of the $f(R) = \left( R^b - \Lambda \right)^c$ gravity}
Next, we will consider another viable $f(R)$ DE model. The $f(R) = \left( R^b - \Lambda \right)^c$ was proposed by Ref.\cite{Amendola:2007nt}. This model has original idea from a generalized $\Lambda$CDM model by parameterizing the power of the Ricci scalar and a whole term of the power of the Ricci scalar with the cosmological constant. This model is viable for the $f(R)$ DE. The model was studied by using the dynamical system method and constrained by data from local gravity and cosmology in the FLRW case \cite{Amendola:2007nt}. More importantly, this model will be the viable $f(R)$ DE with the conditions $c \geq 1$ and $bc\approx 1$ \cite{Amendola:2007nt}. Therefore, it is interesting to extend the study of this model to the anisotropic universe counterpart. The $m$ function of the model can be written in the following form
\begin{eqnarray}
m = \frac{(1-c)r}{c} + b - 1  \,.
\end{eqnarray}
The dynamical system of this model is given as follows:
\begin{eqnarray}\label{autonomous:model2}
\frac{d\Sigma}{d\tau} &=& x_1\Sigma - 3\Sigma + \Sigma\left( 2 - x_3 + \Sigma^2 \right)
+ 2\left( 1- \Sigma^2 -x_1 - x_2 - x_3 - z \right),
\nonumber\\
\frac{d x_1}{d\tau} &=& x_1^2 - 3x_1 - 4x_2 - 2x_3 - z + x_1\left( 2 - x_3 + \Sigma^2 \right),
\nonumber\\
\frac{d x_2}{d\tau} &=& x_1x_2 + \frac{c\,x_1x_2x_3}{c(b-1)x_2 + (1-c)x_3} + 2x_2\left( 2 - x_3 + \Sigma^2 \right),
\nonumber\\
\frac{d x_3}{d\tau}&=& -\frac{c\,x_1x_2x_3}{c(b-1)x_2 + (1-c)x_3} + 2x_3\left( 2 - x_3 + \Sigma^2 \right),
\nonumber\\
\frac{d z}{d\tau}&=& x_1z - 3z +2z\left( 2 - x_3 + \Sigma^2 \right).
\end{eqnarray}
We will discuss the fixed points, their stabilities, the shear and the magnetic evolutions below.

\subsubsection{Fixed points and their stabilities}
As of the study in previous subsection, we will consider the properties of each point in turn. There are 4 physical fixed points in this model. We will organize by two cases: isotropic and anisotropic solutions, and the associated fixed points of this model are given below.
\begin{center}
{\it Isotropic solutions}
\end{center}

\begin{itemize}
\item $P_{1}^{(2)}$\,: de-Sitter fixed point

In this case, we obtain the fixed point:
\begin{eqnarray}
\Sigma = 0,~x_1= 0,~x_2= -1,~x_3= 2,~z= 0.
\end{eqnarray}
Since $w_{{\rm eff}}=-1$, the point $P_{1}^{(2)}$ corresponds to de-Sitter
solutions ($\dot{\Theta}=0$) and has eigenvalues
\begin{equation}
\left\{-4,-3,-3,-\frac{-6+3 (1+b) c+A}{2 (-2+c+b c)},\frac{6-3 (1+b) c+A}{2 (-2+c+b c)}\right\}.\nonumber
\end{equation}
where $A\equiv \sqrt{100-4 (17+25 b) c+\left(9+34 b+25 b^2\right) c^2}$. In this case, it can verify that,
\begin{equation}
a(t) = \exp(\lambda t), \quad \lambda = {\rm arbitrary~constant},\quad M^2=0\,.\label{m1con1}\nonumber
\end{equation}
Notice that the stability conditions of the fixed points satisfy
\begin{equation}
\left(b<-\frac{9}{25}\;\&\;\frac{50}{9+25 b}\leq c<\frac{2}{b}\right) \lor \left(b=-\frac{9}{25}\;\&\;c<-\frac{50}{9}\right)\lor\left(-\frac{9}{25}<b<0\;\&\;\left(c<\frac{2}{b} \lor c\geq \frac{50}{9+25 b}\right)\right)\,,\label{usp21}
\end{equation}
\begin{equation}
\lor\left(b=0\;\&\;c\geq \frac{50}{9}\right)\lor\left(b>0\;\&\;\frac{50}{9+25 b}\leq c<\frac{2}{b}\right)\,.\nonumber
\end{equation}

\item $P_{2}^{(2)}$\,: standard matter-like epoch fixed point

In this case, we obtain the fixed point:
\begin{eqnarray}
&&\Sigma = 0,~x_1= 3-\frac{3}{b c},~x_2= \frac{3-4 b c}{2 b^2 c^2},~x_3= 2-\frac{3}{2 b c},~z= \frac{b c (13-8 b c)-3}{2 b^2 c^2},
\nonumber\\
&\stackrel{b\rightarrow \frac1c}{\Rightarrow}& \Sigma = 0,~x_1= 0,~x_2= -\frac{1}{2},~x_3= \frac{1}{2},~z= 1.
\end{eqnarray}
Employing Eqs.(\ref{dlessweff})-(\ref{at}), we obtain in this case
\begin{equation}
w_{\rm eff} = -1 +\frac{1}{bc},\qquad a(t) = a_0(t-t_0)^{\frac{2bc}{3}},\quad M^2=0\,.\nonumber
\end{equation}
This point $P_{2}^{(2)}$ has eigenvalues
\begin{equation}
\left\{-\frac{3}{2 b c},\frac{3}{c},-1,-\frac{-3+3 b c+B}{4 b c (-1+b c)},-\frac{3-3 b c+B}{4 b c-4 b^2 c^2}\right\}.\nonumber
\end{equation}
where $B\equiv\sqrt{81-498 b c+1025 b^2 c^2-864 b^3 c^3+256 b^4 c^4}$. This fixed point is always saddle.

\item $P_{3}^{(2)}$\,: curvature dominate fixed point

In this case, we obtain the fixed point:
\begin{eqnarray}
&&\Sigma = 0,~x_1= \frac{3}{2 b c-1}-1,~x_2= \frac{6}{1-2 b c}+\frac{1}{b c-1},~x_3= \frac{1}{1-b c}+\frac{3}{2 b c-1}+2,~z= 0,
\nonumber\\
&\stackrel{b\rightarrow \frac1c}{\Rightarrow}& \Sigma = 0,~x_1= 2,~x_2= \text{undefined},~x_3= \text{undefined},~z= 0.
\end{eqnarray}
Employing Eqs.(\ref{dlessweff})-(\ref{at}), we obtain in this case
\begin{equation}
w_{\rm eff} =  -1+\frac{4-2 b c}{3-9 b c+6 b^2 c^2},\qquad a(t) = a_0(t-t_0)^{-\frac{(-1+b c) (-1+2 b c)}{-2+b c}},\quad M^2=0\,.\nonumber
\end{equation}
This point $P_{3}^{(2)}$ has eigenvalues 
\begin{equation}
\left\{-\frac{2 b (-2+b c)}{(-1+b c) (-1+2 b c)},\frac{5-4 b c}{-1+b c},\frac{5-4 b c}{-1+b c},-\frac{2 \left(2-8 b c+5 b^2 c^2\right)}{(-1+b c) (-1+2 b c)},\frac{-3+13 b c-8 b^2 c^2}{1-3 b c+2 b^2 c^2}\right\}.\nonumber
\end{equation}
Regarding the above values, we find for unstable fixed points
\begin{equation} \label{unp23}
b>0\;\&\;\left(\frac{1}{b}<c<\frac{5}{4 b}\right),
\end{equation}
and for stable ones\begin{equation} \label{stap23}
\left(b<0\;\&\;\left(\frac{2}{b}<c<-\frac{1}{16} \sqrt{\frac{73}{b^2}}+\frac{13}{16 b}\lor c>\frac{1}{16} \sqrt{\frac{73}{b^2}}+\frac{13}{16 b}\right)\right)\lor\left(b>0\;\&\;\left(\frac{1}{2 b}<c<\frac{1}{b}\lor c>\frac{2}{b}\right)\right).\nonumber
\end{equation}

\begin{center}
{\it Anisotropic solutions}
\end{center}

\item $P_{4}^{(2)}$\,: Jacob magnetic-like (non-zero magnetic field with matter solution: the Jacobs magnetic field model in Bianchi I \cite{jacob:1969}) with curvature fixed point

In this case, we obtain the fixed point:
\begin{eqnarray}
&& \Sigma = -\frac{2 (b c (5 b c-8)+2)}{b c (7 b c-10)+4},~x_1= -\frac{12 (b c-2) (b c-1)}{b c (7 b c-10)+4},~x_2= -\frac{18 (b c-1) (b c (11 b c-20)+8)}{(b c (7 b c-10)+4)^2},
\nonumber\\
&&\, x_3= \frac{18 b c (b c-1) (b c (11 b c-20)+8)}{(b c (7 b c-10)+4)^2},~z= 0,
\nonumber\\
&\stackrel{b\rightarrow\frac1c}{\Rightarrow}& \Sigma = 2,~x_1= 0,~x_2= 0,~x_3= 0,~z= 0.
\end{eqnarray}
Employing Eqs.(\ref{dlessweff})-(\ref{at}), we obtain in this case
\begin{equation}
w_{\rm eff} = \frac{4+6 b c-7 b^2 c^2}{4-10 b c+7 b^2 c^2},\qquad a(t) = a_0(t-t_0)^{\frac{4-10 b c+7 b^2 c^2}{12-6 b c}}\,,\nonumber
\end{equation}
\begin{equation}
M^2= -\frac{3 \left(16-104 b c+222 b^2 c^2-188 b^3 c^3+55 b^4 c^4\right)}{\left(4-10 b c+7 b^2 c^2\right)^2}\,.\nonumber
\end{equation}
The existence of the magnetic fields is given by
\begin{eqnarray}\label{exist}
\left(b<0\land \left(\frac{4}{5 b}-\frac{1}{5} \sqrt{\frac{6}{b^2}}<c<\frac{10}{11 b}-\frac{2}{11} \sqrt{\frac{3}{b^2}}\lor \frac{1}{11} 2 \sqrt{\frac{3}{b^2}}+\frac{10}{11 b}<c<\frac{1}{5} \sqrt{\frac{6}{b^2}}+\frac{4}{5 b}\right)\right)\nonumber
\nonumber\\
\lor \left(b>0\land \left(\frac{4}{5 b}-\frac{1}{5} \sqrt{\frac{6}{b^2}}<c<\frac{10}{11 b}-\frac{2}{11} \sqrt{\frac{3}{b^2}}\lor \frac{1}{11} 2 \sqrt{\frac{3}{b^2}}+\frac{10}{11 b}<c<\frac{1}{5} \sqrt{\frac{6}{b^2}}+\frac{4}{5 b}\right)\right).\nonumber
\end{eqnarray}
This point $P_{4}^{(2)}$ has eigenvalues
\begin{equation}
\left\{-\frac{12 b (-2+b c)}{4-10 b c+7 b^2 c^2},-\frac{3 \left(8-20 b c+11 b^2 c^2\right)}{4-10 b c+7 b^2 c^2},3P^{-},3P^{+},-\frac{3 \left(4-18 b c+11 b^2 c^2\right)}{4-10 b c+7 b^2 c^2}\right\}.\nonumber
\end{equation}
where
\begin{equation}
P^{\mp}\equiv \frac{\left(128-1088 b c+3984 b^2 c^2-8144 b^3 c^3+10028 b^4 c^4-7428 b^5 c^5+3059 b^6 c^6-539 b^7 c^7\mp Q\right)}{\left(2 (-1+b c) \left(4-10 b c+7 b^2 c^2\right)^3\right)}\nonumber
\end{equation}
and
\begin{equation}
Q\equiv\sqrt{(-1+b c)^2 \left(4-10 b c+7 b^2 c^2\right)^4 \left(320-1984 b c+4128 b^2 c^2-3448 b^3 c^3+1001 b^4 c^4\right)}\,.\nonumber
\end{equation}

\end{itemize}
We turn to discuss the physical fixed points from the autonomous system in the $f(R) = \left( R^b - \Lambda \right)^c$ model. First of all, it is interesting to see all of fixed points in the limits of $c \geq 1$ and $b\rightarrow 1/c$ due to the cosmological viability that pointed out in the literature. At the point $P_2^{(2)}$, there is an existence of the standard matter epoch in this model at the limits of $c \geq 1$ and $bc\approx 1$ but the point $P_3^{(2)}$ is an undefined point in this limit. We do keep the general form of the parameters in this model because the autonomous system suffers from the singularity when the fixed points have $x_2=-x_3$ in the function $1/m$ at the limits of $c \geq 1$ and $b\rightarrow 1/c$. We also classify the physical fixed points by considering the magnitude square of the magnetic fields which must be positive..

\subsubsection{The shear and magnetic fields evolutions}

Here we will see behaviors of the shear and magnetic fields evolutions for the $f(R) = (R^b - \Lambda)^c$ model. In this model, we find the 4 physical fixed points. $P_{4}^{(2)}$ has non-zero shear and magnetic fixed point. Substituting the anisotropic fixed point to the exact solution of the shear evolution from Eq.(\ref{exactshear}), we find,
\begin{eqnarray}
\sigma = \sigma_0 a^{-\eta} = \sigma_0 a_0^{\left(\frac{2 b c - 4}{7 b^2 c^2-10 b c+4}\right)} (t-t_0)^{-\frac{1}{3}}\,,\qquad \eta = \frac{4-2 b c}{7 b^2 c^2-10 b c+4}\,.
\end{eqnarray}
Surprisingly, the shear dissipation in this model has the same cosmic time scale as the $f(R) = R - \alpha R^n$ gravity i.e. $\sigma \sim t^{-\frac13}$. The conditions for the FSD and SSD regions from the $\eta$ parameter are given by
\begin{eqnarray}\label{SSDFSD2}
&& \frac{4-2 b c}{7 b^2 c^2-10 b c+4} > 1,\quad \stackrel{\rm FSD}{\rightarrow} \quad \left(b<0\land \frac{8}{7 b}<c<0\right)\lor \left(b>0\land 0<c<\frac{8}{7 b}\right)\,,
\nonumber\\
&& \frac{4-2 b c}{7 b^2 c^2-10 b c+4} < 1,\quad \stackrel{\rm SSD}{\rightarrow} \quad \left(b<0\land \left(c<\frac{8}{7 b}\lor c>0\right)\right)\lor \left(b>0\land \left(c<0\lor c>\frac{8}{7 b}\right)\right)\,.
\end{eqnarray}
The magnetic fields evolution has the exact solution in terms of cosmic time in the following form:
\begin{eqnarray}
\tilde H = \tilde H_0 a^{-\kappa} = \tilde H_0 a_0^{-\frac{2}{3}\left(1-\frac{4 (b c (5 b c-8)+2)}{b c (7 b c-10)+4}\right)}
(t -t_0)^{\left(\frac{13 b^2 c^2-22 b c+4}{18-9 b c}\right)}\,,\qquad \kappa = \frac{2}{3}\left(1-\frac{4 (b c (5 b c-8)+2)}{b c (7 b c-10)+4}\right).
\end{eqnarray}
As the same procedure in the previous $f(R)$ model, the numeric results of the shear and the magnetic fields evolutions will be given in the section IV. Noting that the the invariant submanifold issues in the phase space of the dynamical system have been so far discussed in details in Ref.\cite{Carloni:2007}. Regarding our chosen (physical) fixed points, they do not admit any singularity or even generate invariant submanifolds.

\section{Cosmological implications}
In this section, we will discuss some relevant cosmological implications of our models. The cosmological implications of magnetic Bianchi I in viable $f(R)$ models of gravity in this present investigation are of great interest to be highlighted. In the following two subsections we discuss the two models: $f(R) = R - \alpha R^n$ and $f(R) = \left( R^b - \Lambda \right)^c$, separately.

\subsection{The $f(R) = R - \alpha R^n$ gravity}

In this model, we study the $f(R) = R - \alpha R^n$ model of gravity in the presence of a uniform magnetic field. We investigate the influence of the primordial magnetic field on the dynamics of the Bianchi I universe. The physical fixed points from the autonomous system in this model provide physical interest. The general conditions for a successful $f(R) = R - \alpha R^n$ model can be summarized as follows:

\begin{itemize}
        \item The point $P_1^{(1)}$ is a stable fixed point when $0 \leq n < 2$. It behaves like a de-Sitter fixed point featuring a late-time de-sitter acceleration. It can also be the saddle if $n > 2$. At late time, the universe in this model can be described by the de-Sitter acceleration solution given by
 \begin{equation}
a(t) = \exp(\lambda t), \quad \lambda = {\rm arbitrary~constant},\quad M^2=0\,.\label{m1con11}
 \end{equation}

	\item The fixed point $P_{2}^{(1)}$ is always saddle point. The standard matter-dominated epoch with the non existence
of the magnetic field might be represented by this fixed point. It is controlled by the following parameters: 
	\begin{equation}
w_{\rm eff} = -1 +\frac1n,\qquad a(t) = a_0(t-t_0)^{\frac{2n}{3}},\quad M^2=0\,.\nonumber
\end{equation}
	We note that this fixed point becomes the standard matter epoch if $n=1$.
	
	\item For the fixed point $P_{3}^{(1)}$ in the $f(R) = R - \alpha R^n$ model, this fixed point might be presented as the beginning of the universe with the curvature-dominated epoch if it is unstable node with the condition in Eq.(\ref{usp13}). In this epoch, the curvature may drive cosmic inflation. However, the fixed point is stable if it is satisfied the condition in Eq.(\ref{stp13}).
		
	\item The fixed point $P_{4}^{(1)}$ is called the Jacobs magnetic-like (with curvature) fixed point stemming from the fact that it has the analogous solution to the Jacobs magnetic field solution in Bianchi I in GR theory \cite{jacob:1969}. This fixed point is always saddle point. Here at this stage the universe is anisotropic with the existence of the magnetic field. Its existence satisfies the condition given in Eq.(\ref{magp14}). Interestingly, this would also be compelling since the universe with a primordial magnetic field is necessary anisotropic.
	
\end{itemize}

\begin{figure}[h]
\begin{center}		
\includegraphics[width=0.7\linewidth]{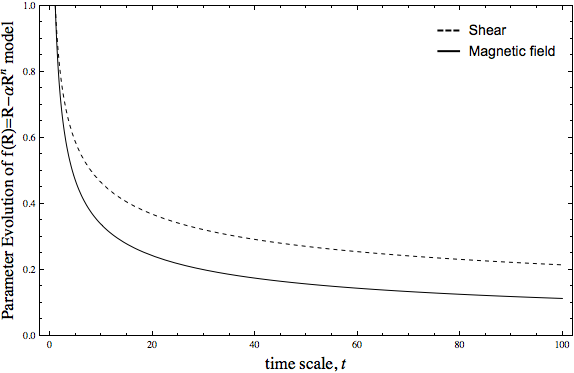}
\caption{\label{smrn}
The plot shows the parameter evolution of the $f(R)=R - \alpha R^{n}$ model. The dashed line shows how shear evolves with the cosmic-time scale; whilst the solid line represents how the magnetic field evolves with the cosmic-time scale for $\sigma_0=a_0=\tilde H_0 = 1$.
}
\end{center}
\end{figure}
Previously, we have already discussed about the shear evolution for all points in the phase space. The conditions of these points to fall whether into the SSD or FSD regions are given in Eq.(\ref{SSDFSD1}). With the given number $n=1.25$ of the parameter in this model, the shear evolution falls into SSD regime when $n > 8/7$. We will end this subsection by examining how shear and magnetic field in this model evolve with time. We find that the shear and the magnetic field will be diluted as illustrated in Fig.(\ref{smrn}). Moreover, by comparing the magnetic field decays a bit faster than the shear.

\subsection{The $f(R) = \left( R^b - \Lambda \right)^c$ gravity}

In this model, we study the $f(R) = \left( R^b - \Lambda \right)^c$ model of gravity in the presence of a uniform magnetic field. We investigate the influence of the magnetic field on the dynamics of the Bianchi I universe. The physical fixed points from the autonomous system in this model provide physical interest. The general conditions for a successful $f(R) = \left( R^b - \Lambda \right)^c$ model can be summarized as follows:

\begin{figure}[h]
\begin{center}		
\includegraphics[width=0.7\linewidth]{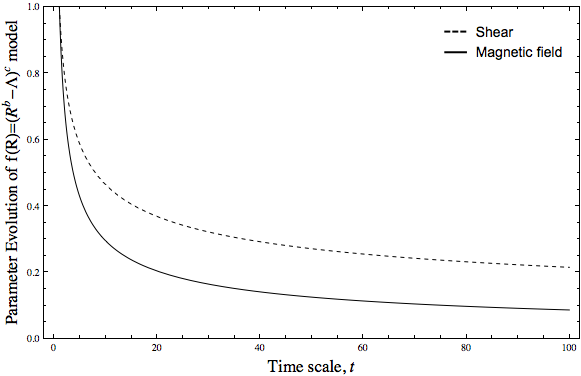}
\caption{\label{smmod}
The plot shows the parameter evolution of the $f(R)=\left(R^{b} - \Lambda\right)^{c}$ model. The dashed line shows how shear evolves with the cosmic-time scale; whilst the solid line represents how the magnetic field evolves with the cosmic-time scale for $\sigma_0=a_0=\tilde H_0 = 1$.
}
\end{center}
\end{figure}

\begin{itemize}

\item The $P_{1}^{(2)}$ point can be the late-time de-Sitter acceleration if it is stable fixed point. Its stable condition of this point is shown in Eq.(\ref{usp21}). On the other hand, this fixed point might be represented as the early epoch of the universe with the condition in Eq.(\ref{usp21}). The saddle point condition is otherwise. The universe at late time can be described by the de-Sitter acceleration parametrized by
	\begin{equation}
	a(t) = \exp(\lambda t), \quad \lambda = {\rm arbitrary~constant},\quad M^2=0\,.\label{m1con12}
	\end{equation}

\item The fixed point $P_2^{(2)}$ is always saddle point. The standard matter-dominated epoch with the non existence of the magnetic field might be represented by this fixed point. It is controlled by the following parameters:
	\begin{equation}
	w_{\rm eff} = -1+\frac{1}{bc},\quad a(t) =a_0(t-t_0)^{\left(\frac{2bc}{3}\right)},\quad M^2=0\,.
	\end{equation}
	Note that this fixed point exactly becomes the standard matter epoch if $bc=1$.

\item For the fixed point $P_3^{(2)}$ in the $f(R) = \left( R^b - \Lambda \right)^c$ model, this fixed point might be presented as the beginning of the universe with the curvature-dominated epoch if it is unstable node with the condition in Eq.(\ref{unp23}). Note that in this epoch the curvature drives cosmic inflation in agreement with the Starobinski model of inflation. To be more concrete, in the following discussion, we will use the specific values of $b$ and $c$ such that $bc \rightarrow 1$ with $c\geq1$. These special values are given by the local-gravity constraints of the viable $f(R)$ DE model for the standard flat-FLRW spacetime (with an isotropic universe) \cite{Amendola:2007nt}. For example, using $b\sim 0.50$, we obtain $2.0<c<2.5$. For our purpose, we select $b\sim 0.50$ and $c \sim 2.33$. Using these values, we come up with only $15\%$ deviation from $bc \rightarrow 1$. However, there are many other choices for their values.
	
\item The fixed point $P_{4}^{(2)}$ is called the Jacobs magnetic-like (with curvature) fixed point, i.e. it has the analogous solution to the Jacobs magnetic field solution in Bianchi I in GR theory \cite{jacob:1969}. This fixed point is always saddle point. Here at this stage the universe is anisotropic with the existence of the magnetic field. Its existence satisfies the condition given in Eq.(\ref{exist}). As of the preceding model, this would also be compelling since the universe with a primordial magnetic field is necessary anisotropic.
	
\end{itemize}

As we already discussed about the shear evolution for all points in the phase space, we then obtain the conditions in which these points fall either into the SSD or FSD regions given in Eq.(\ref{SSDFSD2}). In contrast to the previous $f(R)$ model, the shear evolution of the present model falls into FSD regime with the given number of the parameters $b=0.5$ and $c=2.33$. Here we will furnish this subsection by examining how shear and magnetic field in this model evolve with time. We find that the shear and the magnetic field will be diluted as illustrated in Fig.(\ref{smmod}). Similarly to the preceding model, the magnetic field also decays a bit faster than the shear.

It was noticed that the class of $f(R)$ gravity models based on the isotropic manner which have a viable cosmological expansion chronology, i.e. a matter
dominated epoch followed by a late-time acceleration, was classified in Ref.\cite{Amendola:2007nt}. Here they provided a common value of the parameters of the model for which one can assume the presence of a relevant cosmological orbit. However, in the present investigation, we extended the selected models of $f(R)$ gravity by considering the anisotropic counterpart of flat FLRW metric. Hence, the value of the parameters for the presence of the chronology of a cosmological orbit in our work may deviate from those present in Ref.\cite{Amendola:2007nt}.

\section{Conclusions}
In this work, we study the cosmological dynamics of the magnetic Bianchi I with viable $f(R)$ model of gravity. The dynamical system analysis are utilized to examine the viable $f(R) = R - \alpha R^n$ and $(R^b - \Lambda)^c$ models. In summary, we can highlight our study into 2 distinct cases:
\begin{itemize}
\item For the $f(R) = R - \alpha R^n$ model, we found 4 physical fixed points. There are 3 isotropic solutions and 1 anisotropic case with the presence of primordial cosmological magnetic fields. Based on the viable cosmological sequence, by taking $n=1.25$, the universe starts with the isotropic spacetime with curvature-dominated epoch ($P_3^{(1)}$) and it develops to the anisotropic universe with the presence of the primordial cosmological magnetic fields ($P_4^{(1)}$). After that, the universe isotropizes with the standard matter epoch ($P_2^{(1)}$) and evolves to the de-Sitter late-time acceleration scenario ($P_1^{(1)}$). Eventhough, the given number $n=1.25$ violates the local gravity and cosmological constraints but its constraint is viable only in the FLRW counterpart. Our numerical value, $n=1.25$, might be correct in the magnetic Bianchi I background. The shear evolution of this model has the scale in the cosmic-time as $t^{-\frac13}$ which is slower than the standard GR. The magnetic fields play an important role on the shear dissipation as shown in Eq.(\ref{exactshear}). The primordial cosmological magnetic fields decay a little bit faster than the shear with almost the same scale.

\item The $f(R) = (R^b -\Lambda)^c$ gravity has 4 physical fixed points obtained from the autonomous system. It has the same cosmological chronology as the previous model. The sequence of the universe is $P_3^{(2)}\rightarrow P_{4}^{(2)} \rightarrow P_2^{(2)} \rightarrow P_1^{(2)}$ which gives the reasonable evolution of the universe history with the parameters $b = 0.5$ and $c=2.33$. In addition, these parameter values are compatible with the conditions of the viable $f(R)$ DE of this model, i.e. $c \geq 1$ and $bc\approx 1$. Surprisingly, the $f(R) = (R^b - \Lambda)^c$ model has the same cosmic-time scale of the shear evolution as the previous model, a.k.a. $t^{-\frac13}$.
\end{itemize}

One notes that these 2 models fall into the class A1 of the $f(R)$ model, i.e. they have the de-Sitter stable point at late-time. The explicit treatment of these two models have been carried out in Ref.\cite{Amendola:2007nt} on the FLRW background. Our present study is extended to the anisotropic counterpart of flat FLRW metric. We found that the presence of the anisotropic geometry with LSR from Bianchi I background and the cosmological magnetic fields give an additional fixed point before the emergence of the standard matter epoch. This fixed point shows the existence of the primordial magnetic fields and the anisotropy of spacetime before the universe expands to become the isotropic geometry. The shear evolution modifies dissipative behavior by the primordial cosmological magnetic fields significantly as $\sigma \sim t^{-\frac13}$ (in both of two $f(R)$ models). While for the absence of the magnetic fields case, it gives $\sigma \sim t^{-1}$. In addition, the shear dissipation of the $f(R) = R - \alpha R^n$ gravity is in the SSD regime with the given number $n=1.25$; whilst the shear evolution in the $f(R)= (R^b - \Lambda)^c$ model is in the FSD regime with $b=0.5$ and $c=2.33$.

Based on the viable $f(R)$ DE models, moreover, the reasonable evolution of the universe history for the $f(R) = (R^b - \Lambda)^c$ gravity with $b=0.5$ and $c=2.33$ is more compatible with the viable conditions ($c\geq 1$ and $bc\approx 1$) than the $f(R) = R - \alpha R^n$ model with $n=1.25$ (the viable one is $0<n<1$). It is worth noting that more complicated versions of viable $f(R)$ models (e.g. Starobinski \cite{Starobinsky:2007hu} and Hu-Sawicki \cite{Hu:2007nk} models) have no close forms of the $m$ function written in terms of the variable $r=x_3/x_2$ by using the standard dynamical system approach. However, the authors of Refs.\cite{Carloni:2015jla} proposed a new approach of the dynamical system to handle the problem. Our forthcoming work is to use such the new approach to tackle the Starobinski and Hu-Sawicki $f(R)$ models.

\section*{Acknowledgments}
XL acknowledges support by National Natural Science Foundation of China (Project No. 11547182), and the Doctoral Scientific Research Foundation of Liaoning Province (Project No. 201501197). PC thanks the Institute for the Promotion of Teaching Science and Technology for financial support (grant No. 033/2556). DS acknowledges support from Rajamangala University of Technology Isan, Suranaree University of Technology (SUT) and the Office of the Higher Education Commission under NRU project of Thailand (SUT-COE: High Energy Physics \& Astrophysics) and Thailand Research Fund (TRF) under contract No. MRG5980255.
DS thanks Sante Carloni for reading the manuscript.

\end{document}